\documentclass[iop]{emulateapj}
\slugcomment{{\sc }}
\usepackage{amsmath}
\usepackage{hyperref}
\usepackage[switch]{lineno}
\renewcommand{\abstractname}{Abstract}
\usepackage{perpage}
\MakePerPage{footnote}
\def\apjs{{\rm ApJS}}                  
\def\apjs{{\rm ApJ}}                   
\def\apjl{{\rm ApJ Let}}               
\def\Msol{\thinspace\hbox{$\hbox{M}_{\odot}$}}

\def\a4{\hsize 17.0cm \vsize 25.cm}

\defcitealias{Kroupa2001}{Kroupa}
\defcitealias{MartinezGonzalezetal2018}{I}
\defcitealias{MartinezGonzalezetal2019}{II}
\shorttitle{Mechanical Torque Disruption of Dust Grains Induced by Supernova Shock Waves}
\shortauthors{Martínez-González, S.}

\begin{document}
\slugcomment{{\sc Accepted by ApJ:} March 7th, 2025.}
\title{Mechanical Torque Disruption of Dust Grains Induced by Supernova Shock Waves}
\author{Sergio Martínez-González\altaffilmark{1}}
\altaffiltext{1}{Instituto Nacional de Astrofísica, Óptica y Electrónica, AP 51, 72000 Puebla, México: sergiomtz@inaoep.mx}

\begin{abstract}
The feedback from massive stars drives the evolution of interstellar dust grains by altering their physical properties via a number of radiative and mechanical processes. Through these interactions, interstellar grains can achieve high rotational velocities due to unbalanced torques, potentially leading to their disruption. Mechanical torque disruption occurs when gas–grain collisions, induced by the passage of shocks, spin grains to critical rotational velocities. This study aims to investigate the effects of stochastic mechanical torque disruption on both pre-existent and supernova-condensed dust grains located within wind-blown bubbles. The impact of mechanical torque disruption on supernova-condensed dust and dust grains in wind-blown bubbles is investigated through post-processing of three-dimensional hydrodynamical simulation outputs. The associated timescale is then compared to those of kinetic sputtering and grain shattering. Before the supernova explosion, dust grain disruption timescales within wind-driven bubbles are on the order of millions of years due to the low-density environment. The timescales for mechanical torque disruption (METD) are longer than those for kinetic sputtering and comparable to those of grain shattering, primarily due the high grain drift velocities typical of these regions.
\end{abstract}
  

\section{Introduction} 
\label{sec.intro}

Massive stars shape their environments by the injection of copious amounts of radiation and stellar winds that interact with the surrounding ambient interstellar medium \citep[ISM, e.g.][]{Kourniotisetal2023}. This interaction is governed by the establishment of pressure gradients, which can arise from thermal, dynamical, and magnetic pressures. Massive stars not only shape the dynamics of their surrounding medium but also significantly impact the lifecycle of dust grains within it. These dust grains, which are crucial for various astrophysical processes such as star formation and cooling of the interstellar medium, are subjected to intense radiation and mechanical forces, as well as shocks originating from stellar winds and supernovae. As a result, understanding the mechanisms by which dust grains are disrupted within these hostile environments is crucial to accurately addressing dust evolution in galaxies at all cosmic epochs \citep{schneiderandMaiolino2024}.

Several studies have now considered a non-homogeneous, multiphase interstellar medium \citep[e.g.][]{Priestleyetal2022,Kirchschlager2024,DedikovandVasiliev2025} and the shaping of the interstellar medium by pre-SN massive star activity \citep{MartinezGonzalezetal2018,MartinezGonzalezetal2019,MartinezGonzalezetal2022}, which results in wind-driven shells confining supernova remnants. Important outcomes for the fate of pre-existing and ejecta-condensed dust grains affected by pre-supernova ISM-shaping are summarized as follows:

\begin{enumerate}
    \item {CSM dust grains}. In scenarios involving stars that experienced massive eruptions years before the final explosion, a significant fraction of the supernova kinetic energy (up to half of it) is radiated away upon SN shock-CSM interaction. As a result, both the SN forward and reverse shocks are weakened. This weakening not only helps to preserve a large fraction (between $20$--$75\%$) of the CSM dust mass, but is also expected to reduce the destruction of pre-existing ISM dust grains and ejecta-condensed dust grains \citep{SerranoHernandezetal2025}.
    
    \item {Pre-existent ISM dust grains}. An encompassing massive wind-driven shell (WDS) provides a protective barrier for interstellar dust grains, sweeping and stacking them within the WDS long before the SN explosion. Soon after the explosion, when the SN forward shock wave reaches the WDS, radiative cooling readily weakens the transmitted blast wave (while being partially reflected), which only traverses a small layer within the WDS \citep[][]{TenorioTagleetal1990,Dwarkadas2007,vanMarleetal2015,Haidetal2016}. Thus, only a small amount, a fraction of a solar mass, of preexistent interstellar dust grains would be destroyed \citep{MartinezGonzalezetal2019,MartinezGonzalezetal2022}.

    \item {Ejecta-condensed dust grains}. The SN remnant rapidly expands within the tenuous region excavated by the stellar wind. As a consequence, gas-grain and grain-grain collisions are infrequent in such a region. This low-density condition continues to be valid during the crossing of the reverse shock into the supernova ejecta. Thus, a significant fraction of the dust generated within the supernova ejecta would remain largely unaffected by thermal sputtering
\citep{MartinezGonzalezetal2018,MartinezGonzalezetal2019,MartinezGonzalezetal2022}.

\end{enumerate}

However, a number of studies have incorporated important ingredients that are missing in the previous analysis \citep[][and references therein]{Bocchioetal2016,Kirchschlageretal2023,Fryetal2020,Slavin2024}. Among these ingredients, the disruption of spinning dust grains due to mechanical stress has been presented by \citet{Purcell1975,DraineandSalpeter1979b,HoangandLee2020} and references therein. In this context, the unbalanced torques acting on dust grains can cause them to spin at high angular speeds. As their rotation speeds increase, they may reach a critical point where the centrifugal stress on each grain exceeds its maximum tensile strength. At this moment, the grains begin to break apart into tiny smaller particles. In the following sections, the impact of mechanical torque disruption (METD) within wind-driven bubbles is evaluated in comparison to kinetic sputtering and grain-grain collisions leading to grain shattering.

The structure of the paper is as follows: Section \ref{sec:model} presents the setup and initial conditions of the hydrodynamical models used to simulate the wind-blown bubble and supernova explosion. Section \ref{sec:post-processing} describes the post-processing analysis, including the mechanical torque disruption mechanism and other dust destruction processes such as kinetic sputtering and grain shattering. The results of these simulations, focusing on the characteristic timescales and physical interactions in the medium, are presented and discussed in Section \ref{sec:results}. Finally, Section \ref{sec:discussion} discusses the main findings and their implications, while Section \ref{sec:conclusions} summarizes the main conclusions.

 \section{Hydrodynamical Model}
\label{sec:model}

\begin{figure*}
    \centering
    \includegraphics[width=1.0\textwidth]{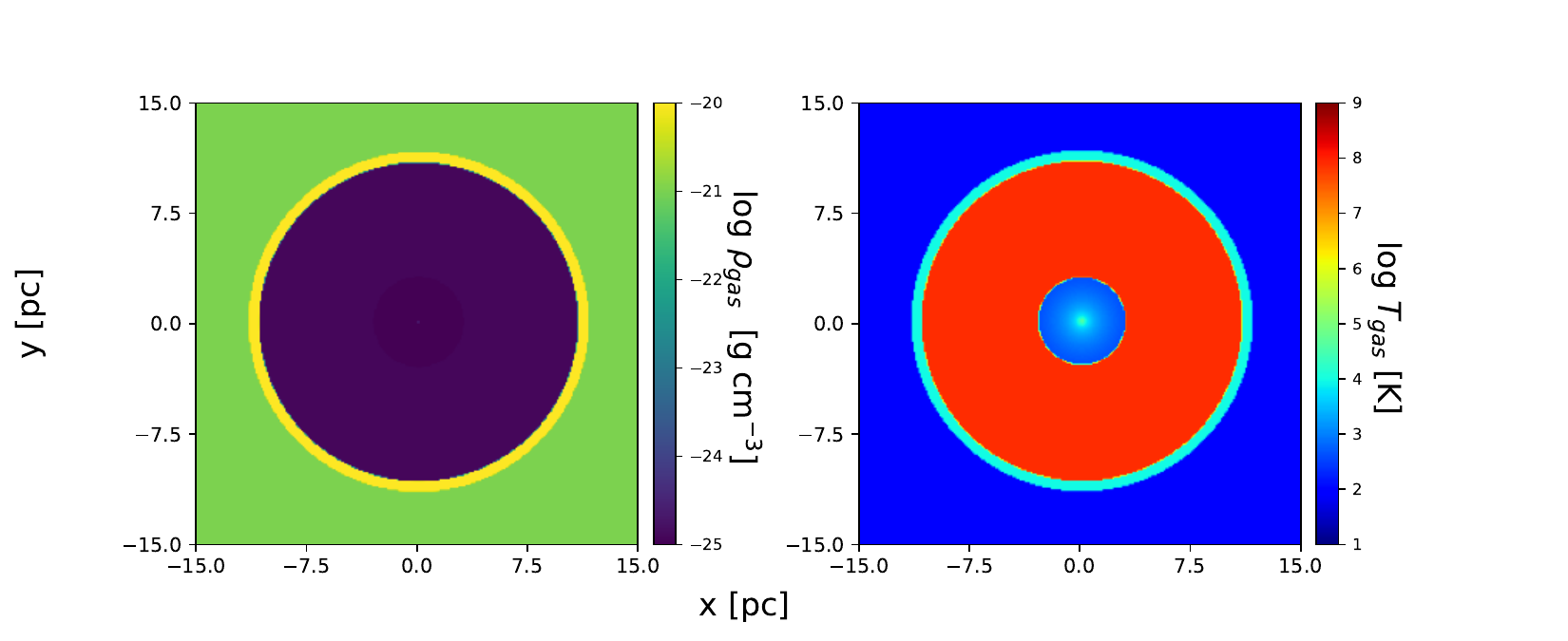}
    \caption{Spherically symmetric wind structure from a $60$-M$_\odot$ massive star mapped onto a three-dimensional Cartesian grid. The left panel displays the logarithm of the gas mass density, $\log(\rho_{gas}/\text{g cm}^{-3})$, while the right panel shows the logarithm of the gas temperature, $\log(T_{gas}/\text{K})$. See \citep{MartinezGonzalezetal2019} for further details.}
    \label{fig:0}
\end{figure*}

An instance of a three-dimensional hydrodynamical simulation of a wind-blown bubble and the subsequent supernova explosion from the progenitor massive star serves as an illustrative approach to evaluate the characteristic timescales for dust grain disruption. The FLASH 4.6.2 adaptive mesh refinement (AMR) code \citep{Fryxelletal2000} is employed for this purpose. FLASH applies the Piecewise Parabolic Method (PPM) to solve the hydrodynamic equations \citep{ColellaandWoodward1984} within an adaptive mesh refinement framework. For simplicity, the simulation framework incorporates a time-dependent spherically-symmetric stellar wind modeled through the Wind module \citep[]{Wunschetal2017}. The initial medium in which the wind-blown bubble grows has a density of $10^3$ cm$^{-3}$. The wind structure exhibits a stratified configuration as described by \citep{Weaveretal1977}: a free-wind region near the star, where the stellar wind moves without significant interaction with the surrounding medium; a shocked-wind region, where the stellar wind has decelerated; a wind-driven shell (WDS), created by the piling-up of ambient gas as it is swept by the wind's leading shock, which is separated from the shocked-wind region by a contact discontinuity; and the undisturbed ambient medium beyond the shell, which remains unaffected by the wind (see Figure \ref{fig:0}).

Once the progenitor star, with zero-age mass 60 M$_\odot$ and formed out of solar metallicity gas, has reached an age of $\sim 3.8$ Myr and is about to explode as a supernova, the one-dimensional spherically symmetric wind structure is then mapped onto a three-dimensional Cartesian grid spanning a volume of $15 \times 15 \times 15$ pc$^3$. The grid is resolved with $256$ cells along each spatial dimension, providing high resolution for subsequent analysis. 

The supernova explosion releases $10\,\Msol$ of ejecta mass ($M_{ej}$) and $E_{SN} = 10^{51}$\,erg in kinetic energy. The ejecta mass density distribution is given by \citep{TrueloveandMckee1999,TangChevalier2017}

\begin{equation}
\rho_{ej} = \frac{(3 - \omega)}{4\pi} \frac{M_{ej}}{R_{SN}^3} \left( \frac{R_{SN}}{r} \right)^\omega,
\end{equation}

where $r$ is the radial distance from the center of the explosion and $\omega = 1$ is adopted. The velocity distribution of the ejecta is described by:

\begin{equation}
v_{ej} = \left( 2 \frac{(5 - \omega)}{(3 - \omega)} \frac{E_{SN}}{M_{ej}} \right)^{1/2} \left( \frac{r}{R_{SN}} \right).
\end{equation}

The explosion occurs at the center of the computational domain with $R_{SN}=0.5$ pc. This simulation is analogous to that presented by \citet{MartinezGonzalezetal2019} at the same density and stellar wind parameters, except that now the effects of an initially perturbed ambient velocity field with a small dispersion velocity of $1$ km s$^{-1}$ is included. The supernova ejecta is inserted with a temperature of $10^7$ K in order to mimic the effect of heating by the radioactive decay chain ${}^{56}\textrm{Ni}\rightarrow{}^{56}\textrm{Co}\rightarrow{}^{56}\textrm{Fe}$ \citep[see][]{SerranoHernandezetal2025}.

At the crossing of a supernova forward/reverse/reflected shock, gas particles and the population of dust particles are expected to acquire different velocities, which triggers not only the METD mechanism, but also the onset of kinetic sputtering and grain shattering by grain-grain collisions \citep[e.g.][]{Bocchioetal2016}. The current hydrodynamical implementation, presented in \citep{MartinezGonzalezetal2018,MartinezGonzalezetal2019,MartinezGonzalezetal2022}, does not account for the decoupled motion of gas and dust particles, which is crucial for accurately modeling the interaction between the two components. Therefore, a set of post-processing routines is applied to the output of the hydrodynamical simulations to quantify the role of the mechanisms triggered by this decoupled motion. These routines are applied only to the gas that has already undergone shock processing.

 \section{Post-processing}
\label{sec:post-processing}
\subsection{Mechanical torque disruption}
\label{subsec:metd}

\begin{figure*}
    \centering
    \includegraphics[width=0.9\textwidth]{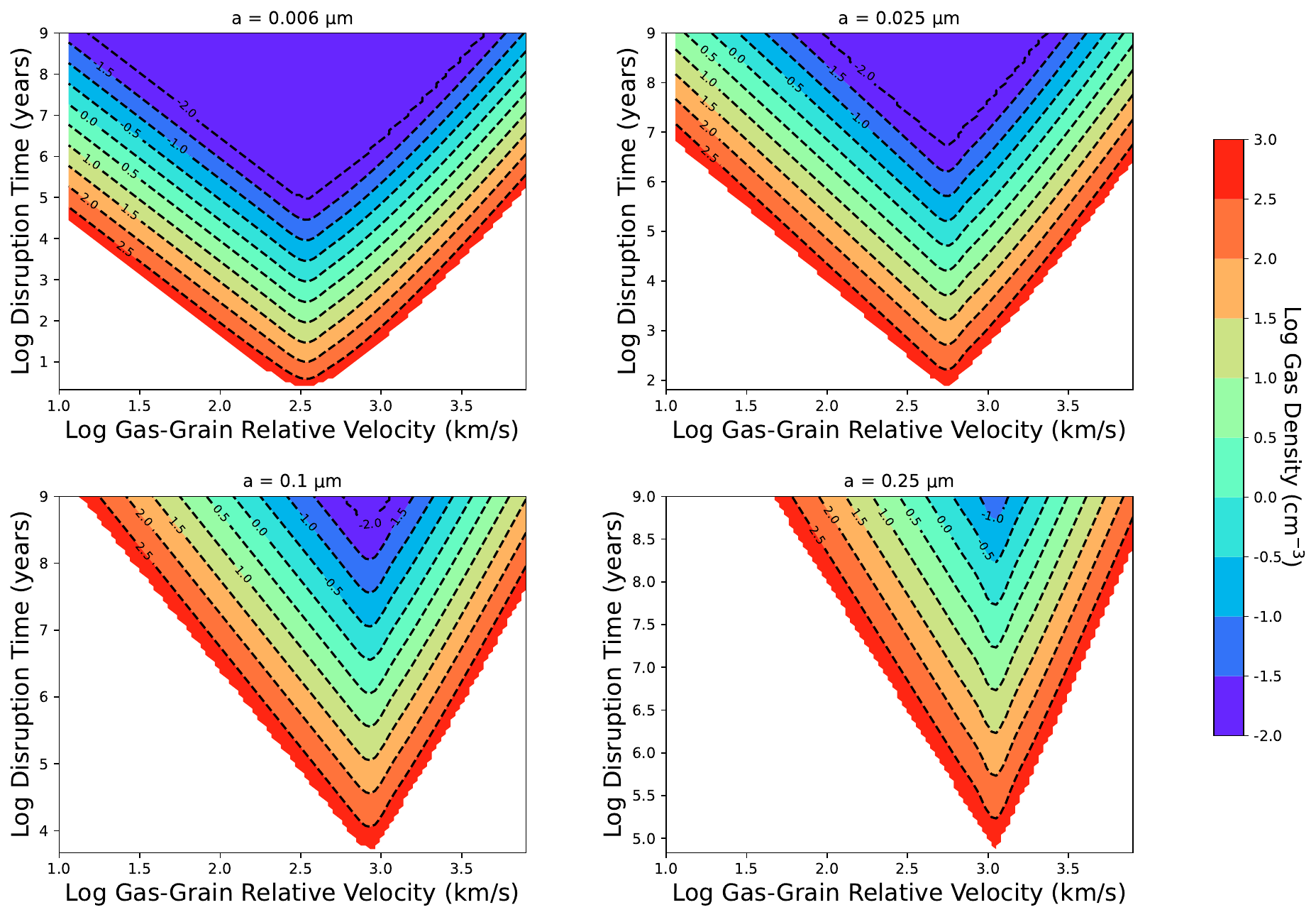}
    \caption{The logarithm of the disruption time, ($\log \tau_{\mathrm{METD}}$ [yr]), is presented as a function of the logarithm of the relative gas-grain velocity, $(\log v_{d}$ [km s$^{-1}$]). Panels correspond to dust grain radii $a = 0.006$, $0.025$, $0.1$, and $0.25~\mu$m, illustrating how disruption time increases with decreasing gas densities. The color bar indicates the gas number density, $(\log n$ [cm$^{-3}]$), with dashed lines marking specific values, which are labeled above each line. This figure is analogous to Figure 2 of \citet{HoangandLee2020} but for a range of gas densities.}
    \label{fig:1}
\end{figure*}

\citet{HoangandLee2020} developed a formalism for stochastic mechanical torque disruption (METD), where gas-grain collisions spin up dust grains until centrifugal stress shatters them. They found that METD is more efficient than non-thermal sputtering for sub-10 nm grains at velocities below 500 km s$^{-1}$. The disruption time for this process, $\tau_{METD}$, can be described by the equation,

\begin{eqnarray}
\tau_{METD} = \left(\frac{2 I^2 \omega_{cri}^{2}}{ m_H \rho_{gas} v_{d}^{3} \pi a^{4}} \right) \left(\frac{1}{f_{p}}\right)^{2},
\end{eqnarray}

where $a$ is the grain radius, $I=8 \pi \rho_{gr} a^5 / 15 $ is the grain inertial momentum, $\rho_{gr}$ is the bulk density of the dust grain, $\omega_{cri}$ is the critical angular velocity for grain disruption, $m_H$ is the hydrogen mass, and $f_{p}$ is the efficiency of gas-grain momentum transfer. The critical angular velocity for grain disruption is given by

\begin{eqnarray}
\omega_{cri}=\frac{2}{a}\left(\frac{S_{max}}{\rho_{gr}} \right)^{1/2}~,
\label{eq:omega_cri}
\end{eqnarray}

where $S_{max}$ is the tensile strength of the grain material \citep[see][for further details]{HoangandLee2020}. $S_{\max}$ can be estimated by the analytical formula \citep{Tatsuumaetal2019,Hoangetal2021},

\begin{eqnarray}
S_{\max} &\simeq& 9.51 \times 10^8 \left(\frac{\gamma_{\text{sf}}}{1000 \text{ erg cm}^{-2}}\right) \left(\frac{r_0}{0.1 \ \text{nm}}\right)^{-1} \nonumber \\
&\times&
\left(\frac{\phi}{0.1}\right)^{1.8} \text{ dyn cm}^{-2},
\label{eq:smax}
\end{eqnarray}

for carbonaceous grains. Taking the monomer radius as $r_0 = 0.1281$ nm, the surface energy as $\gamma_{\text{sf}} = 1400$ erg cm$^{-2}$ \citep{TodiniandFerrara2001}, and the grain volume filling factor as $\phi = 0.1$ \citep{Hoangetal2021}, one thus obtains

\begin{equation}
S_{\max} \simeq 1.04 \times 10^9 \text{ dyn cm}^{-2}.
\end{equation}

Thus, a maximum tensile strength $S_{\max} = 10^9$ dyn cm$^{-2}$ is adopted across all grain sizes (although $\tau_{\text{METD}} \propto S_{\max}$ allows for straightforward scaling for different tensile strengths). Note
that the conservative value of $\phi = 0.1$ corresponds to porous structures, while dust grains in hot gas likely have a compact structure resulting in a higher tensile strength \citep{HoangandLee2020}.

A grid of $10^4$ METD models has been calculated following the aforementioned formalism to determine $\tau_{\text{METD}}$.
This grid (see Figure \ref{fig:1}) provides a comprehensive set of $\tau_{METD}$ values across a range of grain and gas properties, allowing for interpolation during the post-processing of the hydrodynamical results.

The METD grid covers a wide range of relative gas-grain velocities from 10 to 10$^4$ km s$^{-1}$ and spans gas number densities from $n = 10^{-3}$ to 10$^3$ cm$^{-3}$. As shown, there is a strong dependence of mechanical torque disruption on both grain size and gas-grain relative velocity.

By interpolating within each grid cell, one can determine the $\tau_{METD}$ values corresponding to the simulated hydrodynamical conditions. 

\begin{figure*}
    \centering
    \includegraphics[width=0.9\textwidth]{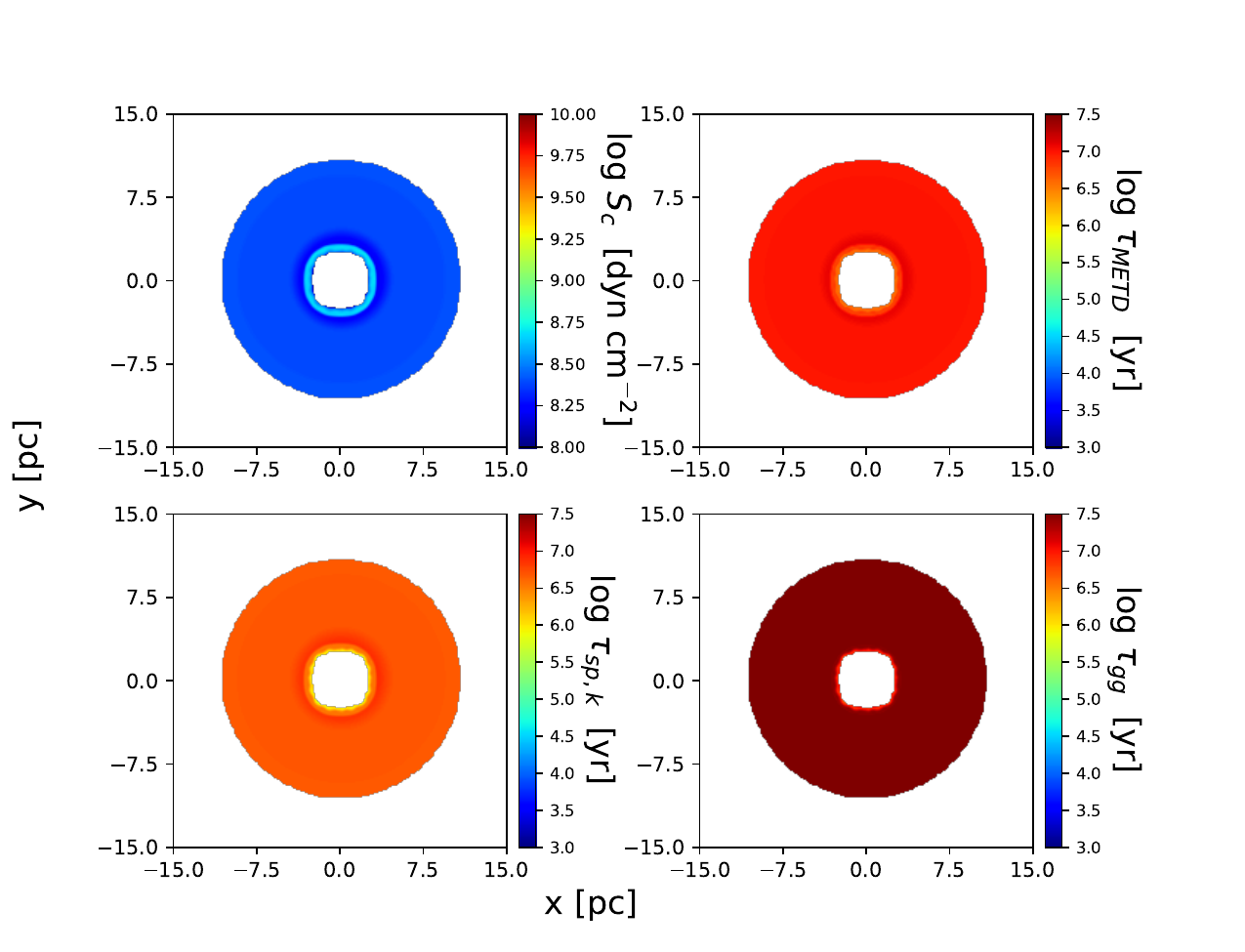}
\caption{Density and timescale maps for the simulation corresponding to $\sim1$ kyr post-explosion. From top left to bottom right: centrifugal stress ($\log S_c$ [dyn cm$^{-2}$]), METD disruption timescale ($\log \tau_{\mathrm{METD}}$ [yr]), kinetic sputtering timescale ($\log \tau_{\mathrm{sp}}$ [yr]), and grain shattering timescale ($\log \tau_{\mathrm{gg}}$ [yr]). The inner white region in the upper right and lower panels, which appears unfilled, actually corresponds to the location of unshocked gas, where the motion of gas and dust particles is assumed to be tightly coupled.}
    \label{fig:timescales}
\end{figure*}

\begin{figure*}
\centering
\includegraphics[width=0.9\textwidth]{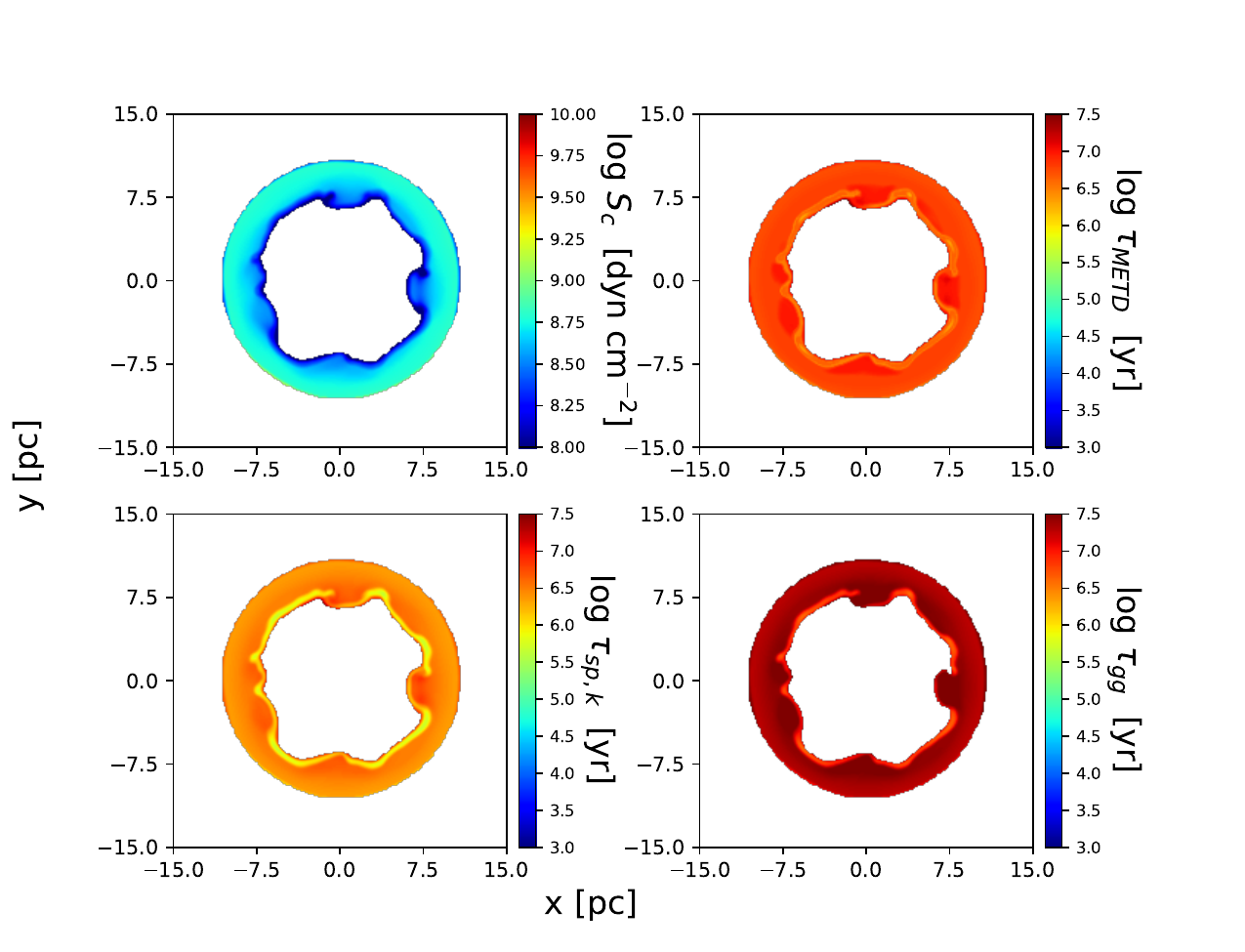}
\caption{Same as Figure \ref{fig:timescales} but showing a snapshot at $\sim3.9$ kyr post-explosion. The supernova blast wave is about to collide with the encompassing wind-driven shell. As in Figure \ref{fig:timescales}, the inner white region in the upper and lower panels corresponds to the location of the unshocked supernova ejecta.}
\label{fig:timescales2}
\end{figure*}

\begin{figure*}[htbp]
\centering
\includegraphics[width=0.9\textwidth]{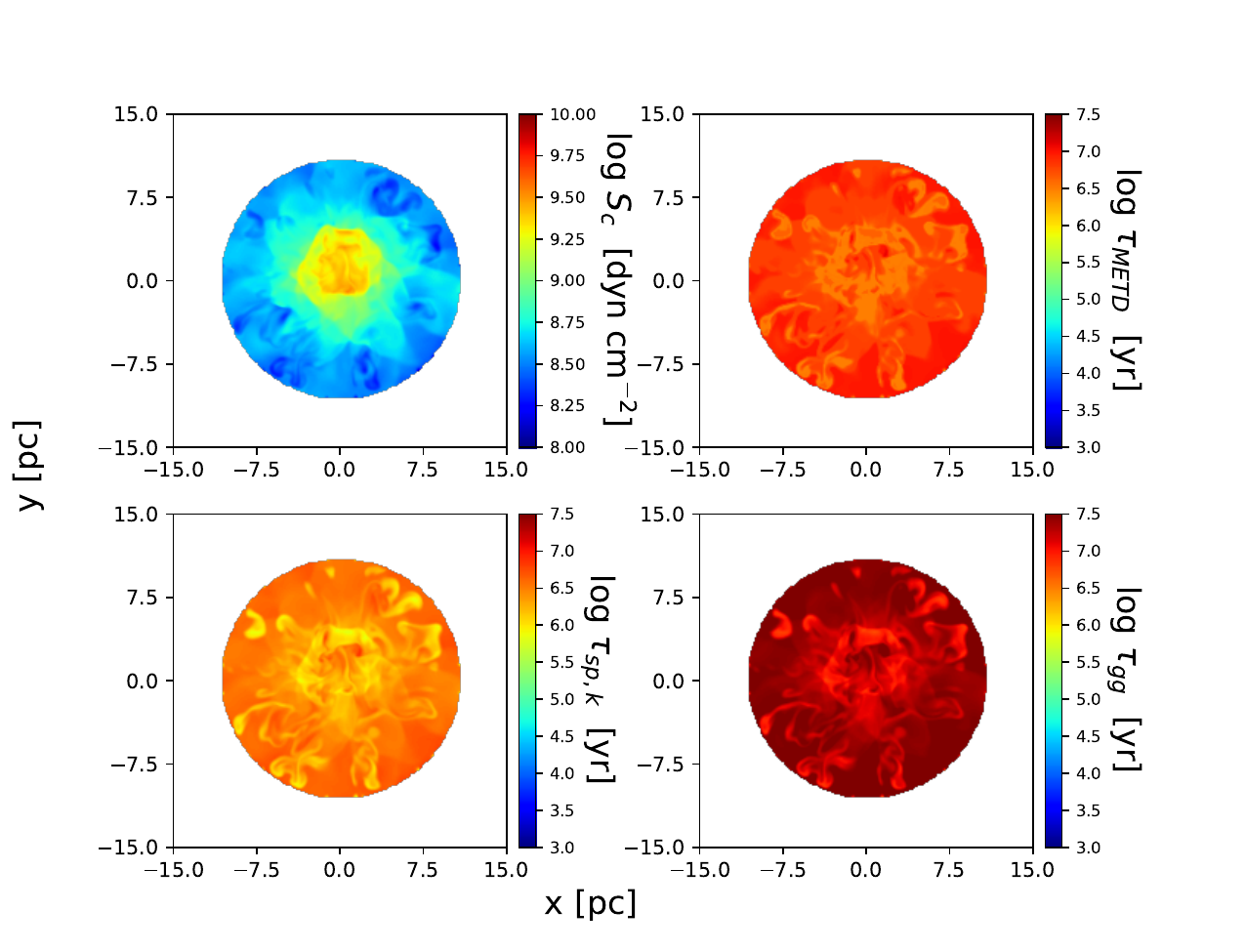}
\caption{Same as Figure \ref{fig:timescales} but showing a snapshot at $\sim7.9$ kyr post-explosion. The supernova blast wave has being partially reflected after the collision and has reached the explosion site to start reverberating. Note the redder region in the upper left panel, which begins to advance forward with increased centrifugal stress by more frequent gas-grain thermal collisions.}
\label{fig:timescales3}
\end{figure*}

\subsection{Rotational disruption by thermal collisions}

Dust grains spinning as a result of thermal collisions within a hot plasma can also be torn apart at sufficiently large temperatures. The centrifugal stress ($S_c$) experienced by a rotating dust grain in a hot plasma, considering energy equipartition and radiative damping, can be expressed as \citep{DraineandSalpeter1979b}

\begin{eqnarray} S_c &=& 2.9 \times 10^9  \text{dyn cm}^{-2} \left( \frac{a}{0.01 \mu\text{m}} \right)^{-1/2} \left( \frac{\epsilon U}{\text{V}} \right)^{-1}
\nonumber \\
&\times&
\left( \frac{\rho_{gr}}{\text{g cm}^{-3}} \right)^{1/2} \left( \frac{n}{\text{cm}^{-3}} \right)^{1/2} \left( \frac{T}{10^8 \text{K}} \right)^{3/4}. \end{eqnarray}

where $n$ is the gas number density. Gas-grain thermal collisions can be an important route for grain disruption for shocked gas within wind-blown bubbles. Thus $S_c$ has also been calculated in order to compare it to the maximum tensile strength $S_{max}$ using $\epsilon$=0.05 and a grain electrostatic potential $U=100$ V.

\subsection{Kinetic sputtering and grain shattering}
\label{subsec:kin-shat}

The timescales for kinetic sputtering and grain shattering are \citep[e.g. ][]{HoangandTram2019}:

\begin{eqnarray}
\tau_{sp,k} = \frac{4a\rho_{gr}}{\rho_{gas} v_{d} Y_{sp,k}}, & \quad & \tau_{gg} = \frac{4a\rho_{gr}}{3\rho_{gas} \delta v \mathcal{D}}.
\end{eqnarray}

where $\rho_{gas}$ is the gas mass density, $Y_{sp,k}$ denotes the kinetic sputtering yield, $v_{d}$ indicates the relative velocity between the gas particles and grains, $\delta v$ defines the relative velocity of colliding dust grains, and $\mathcal{D}$ is the dust-to-gas mass ratio. In this study, spherical carbonaceous grains of radius $a$ are considered to move through a gas medium with a normal chemical composition, containing 1 helium atom per 10 hydrogen atoms.

For the stochastic METD disruption mechanism, a grain size of $0.01$ 
$\mu$m is adopted, as the METD mechanism is more efficient for grains of this size and smaller, as shown by \citet{HoangandLee2020}. This mechanism becomes especially efficient for gas-grain relative velocities $v_{d} \sim 500$ km s$^{-1}$. Using these parameters, the effect of the METD mechanism is maximized.

Similarly, carbonaceous grains with $a = 0.01$ $\mu$m were adopted to evaluate the timescales for kinetic sputtering and grain shattering. The kinetic sputtering timescale, $\tau_{\mathrm{sp}}$, was calculated using the standard values $\rho_{gr}=2.26$ g cm$^{-3}$, $Y_{sp,k}=0.1$, and $v_{d}=500$ km s$^{-1}$. Similarly, the grain shattering timescale, $\tau_{\mathrm{gg}}$, was calculated by assuming $\delta v=20$ km s$^{-1}$ and a dust-to-gas mass ratio $\mathcal{D}=0.01$.

 \section{Results}
\label{sec:results}

Before the supernova explosion, the timescales of grain destruction/disruption within wind-driven bubble are fairly long ($\gtrsim 1$ Myr) owing to the evacuation of the interstellar gas around the eventually exploding star by the stellar wind, leading to low densities $\lesssim 10^{-2}$ cm$^{-3}$. Figure \ref{fig:timescales} shows the post-processed maps for the centrifugal stress experienced by the dust grains immersed in the hot gas, the METD disruption timescale, kinetic sputtering timescale, and grain shattering timescale at $1$ kyr post-explosion. At this moment, the small supernova remnant has expanded to a radius of $\sim 3$ pc. The supernova remnant reaches a radius of $\sim 7$ pc after evolving $3.9$ kyr (see \ref{fig:timescales2}. At later times, the collision between the supernova remnant and the wind-driven shell has resulted in strong radiative cooling. This strong radiative losses thereby inhibit the blast wave's ability to penetrate through the dense wind-driven shell, effectively confining it within the bubble's extent \citep[][and references therein]{MartinezGonzalezetal2019}. At $7.9$ kyr post-explosion (Figure \ref{fig:timescales3}), the reflected blast wave has merged with the supernova reverse shock, reaching the explosion site and subsequently reverberating within the wind cavity. This oscillatory motion further amplifies the centrifugal stress from thermal gas-grain collisions, exceeding the fiducial $S_{max}$. At lower ambient gas densities, however, this reverberation can be expected to be weaker due to the larger volume and lower densities within the wind cavity. 

Notably, at all times, all calculated timescales are extremely long within the region filled with shocked gas, ranging from millions to hundreds of millions of years. This is primarily due to the low-density environment created by the stellar wind, where gas-grain and grain-grain collisions are infrequent. 

\section{Discussion}
\label{sec:discussion}

The results presented by \citet{HoangandLee2020} demonstrate that the destruction of small grains via METD in fast shocks of supernova remnants could be less efficient than that predicted by kinetic sputtering at large gas-grain relative speeds ($v_{d} > 300$ km s$^{-1}$). These findings are consistent with the results presented here, showing that the timescales for METD are longer than those for kinetic sputtering and are comparable to the timescales for grain shattering, particularly for small grains with a size of $a=0.01$ $\mu$m. Moreover, the METD disruption timescale remains fairly long, spanning millions to tens of millions of years. This prolonged timescale is consistent with the high grain velocities and low-frequency collisions characteristic of the hot, low-density gas shaped by the pre-supernova massive star feedback. Throughout the simulations, the grain's maximum tensile strengths were set to $S_{\text{max}} = 10^9$ dyn cm$^{-2}$, which implies that the stochastic METD disruption timescale could be an order of magnitude shorter (longer) for every order of magnitude decrease (increase) in the tensile strength. These prolonged timescales suggest that while stochastic METD could be significant in certain contexts, it is unlikely to dominate grain destruction in the hot gas regions of supernova remnants, where other processes may play a more prominent role in dust grain evolution. The presented results do not account for the role of mechanical torques acting on irregular drifting grains or the effect of radiative mechanical torques (see Appendices \ref{app:rMETD} and \ref{app:RATD} for analytical estimates of these processes).

\section{Concluding Remarks}
\label{sec:conclusions} 

This work has investigated the role of stochastic mechanical torque disruption, as developed by \citet{HoangandLee2020}, in the evolution of dust grains within supernova remnants and wind-blown bubbles. The stochastic METD timescales, along with kinetic sputtering and grain shattering timescales, are computed, revealing that these processes are generally inefficient in the low-density, high-temperature environments characteristic of these regions.

Before the supernova explosion, dust grains within the wind-blown bubble survive for timescales of $\gtrsim 1$ Myr. Post-explosion, and as the remnant expands, radiative cooling confines the blast wave within the wind-driven shell. This later produces reverberations of the blast wave (now a reflected shock) that amplify the centrifugal stress, exceeding the fiducial tensile strength in localized regions, although this effect also weakens in lower-density regions.

Although METD is unlikely to dominate grain destruction in such conditions unless we consider weak materials with low tensile strengths, its influence on the grain size distribution may contribute to shaping the interstellar dust population over long timescales. The results remark on the need to consider multiple processes simultaneously when modeling dust evolution in and around supernova remnants.  

Future studies should explore the interplay between METD and other mechanisms under varying environmental conditions, including the effects of magnetic field strength and orientation. Furthermore, improved modeling of grain properties, such as tensile strength, shape, electric charge, and composition, will provide deeper insight into the survival of dust grains before and after a supernova explosion.

\section{Acknowledgments}

The author thanks the anonymous referee for helpful comments and suggestions, which improved the quality of the paper, and expresses gratitude to Santiago Jiménez Villarraga for engaging and insightful discussions that contributed to a deeper understanding of this topic. The authors thankfully acknowledges computer resources, technical advice and support provided by Laboratorio Nacional de Supercómputo del Sureste de México (LNS), a member of the SECIHTI national laboratories, with project No. 202501004C. {\it Software:} FLASH v4.6.2 \citep{Fryxelletal2000}, Numpy \citet{Numpy}, Wind \citep{Wunschetal2017}, \textsc{Cinder} \citep{MartinezGonzalezetal2018}, Matplotlib \citep{Matplotlib}, SciPy \citep{SciPy}, h5py \citep{collette_python_hdf5_2014}.

\appendix
\section{Dust Grains with Irregular Shapes}
\label{app:rMETD}

Mechanical torques could also be induced on irregular grains by their relative drift to the gas as they reach high angular momentum \citep[commonly referred to as regular mechanical torques, rMET,][]{HoangandTram2022}. In this Appendix, the maximum angular velocity of irregular dust grains spun-up by rMETs is compared to the critical angular velocity for grain disruption (equation \ref{eq:omega_cri}). The maximum angular velocity of irregular grains spun-up by rMETs is given by
\citep{HoangandTram2022}

\begin{equation}
    \Omega_{\rm rMET} = 4.5 \times 10^6 \left( \frac{v_{d}}{v_T} \right)^2 \frac{s T_{\rm gas,1}^{1/2} Q_{\rm spinup,-3}}{a_{-5} \Gamma_{\parallel}} \quad \text{rad s}^{-1},
\end{equation}

where $v_T = \sqrt{2k_B T_{\rm gas}/m_H}$ is the thermal speed of the gas, $s$ is the grain's axial ratio, $T_{\rm gas,1}$ is the gas temperature normalized to 10 K, $Q_{\rm spinup,-3} = Q_{\rm spinup}/10^{-3}$ represents the normalized spin-up efficiency, $a_{-5} = a / 10^{-5} \text{cm}$, and $\Gamma_{\parallel}$ is the geometrical factor which is taken as 1.

\begin{figure*}
    \centering
    \includegraphics[width=0.9\textwidth]{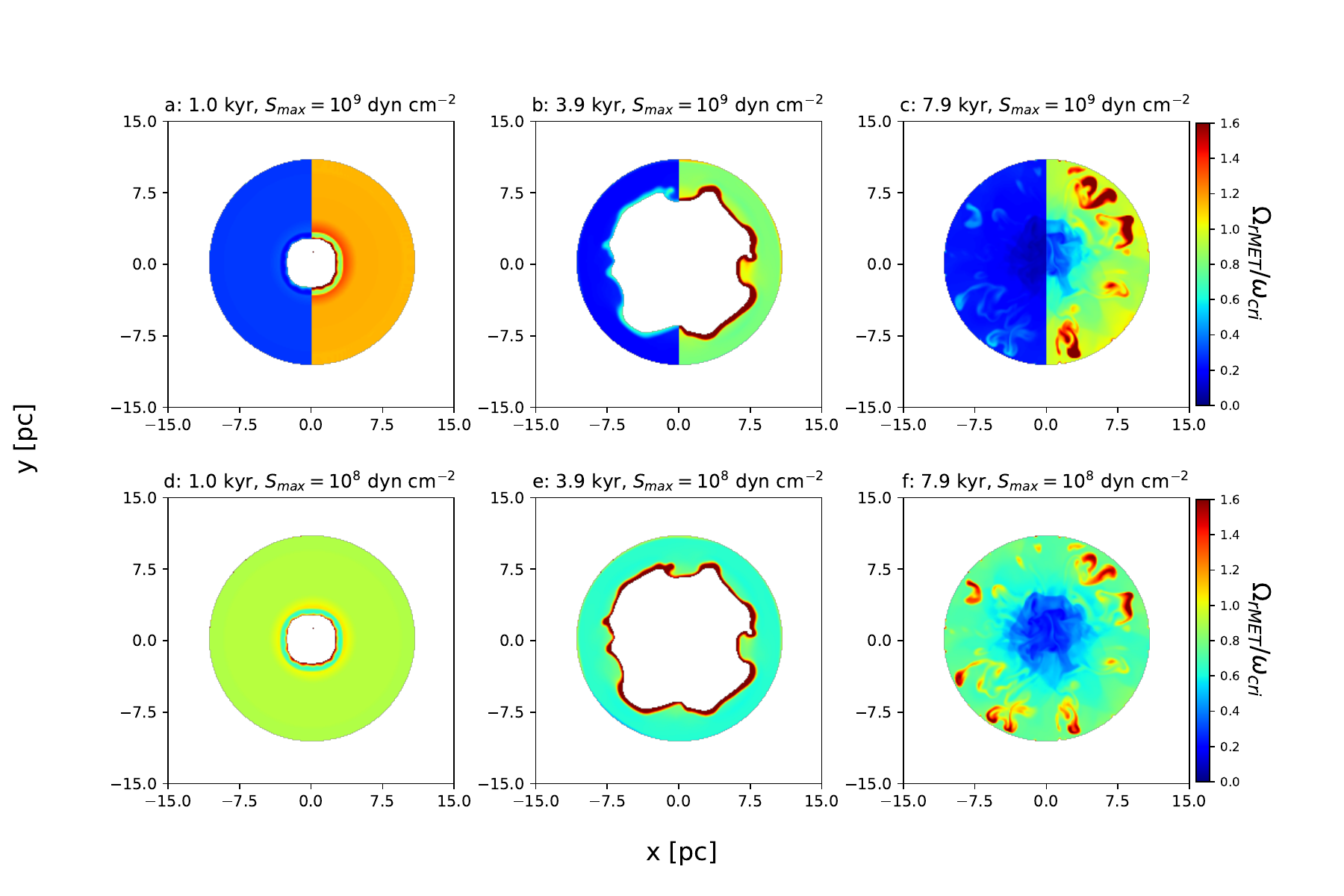}
\caption{Comparison of the ratio between the maximum angular velocity of irregular grains induced by regular mechanical torques (rMETs) and the critical angular velocity $\omega_{\rm cri}$ at various evolutionary times (as indicated on top of each panel). The upper panels show results for $a = 0.01 \ \mu$m and $S_{\rm max} = 10^9$ dyn cm$^{-2}$, where rMETs are largely suppressed in high-temperature regions. In the upper panels, results were computed for $v_d = 500$ km s$^{-1}$ on the left side and $v_d = 1000$ km s$^{-1}$ on the right side, as indicated by the separation in the color scale. The lower panels assume $S_{\rm max} = 10^8$ dyn cm$^{-2}$ and $v_d = 500$ km s$^{-1}$. In both the right side of the upper panels and the lower panels, $\Omega_{rMET}$ becomes larger than $\omega_{cri}$ during the crossing of the reflected shock front (panels b and e) and in lower temperature clumps/filaments after the reverberation of the reflected shock (panels c and f)}.
    \label{fig:A1}
\end{figure*}

Figure \ref{fig:A1} presents the ratio between the angular velocities induced by rMETs and $\omega_{cri}$ at the same evolutionary times as depicted in Figures \ref{fig:timescales}-\ref{fig:timescales3} for
$Q_{\rm spinup}=10^{-3}$ \citep{Hoangetal2018a}, $a=0.01$ $\mu$m and $S_{max}=10^9$ dyn cm$^{-2}$ (upper panels). Assuming $v_d = 500$ km s$^{-1}$ (left side, panels a-c), $\Omega_{\rm rMET}/\omega_{cri}$ remains below unity everywhere, although the ratio increases at the reflected shock front. For $v_d = 1000$ km s$^{-1}$, it exceeds unity across the shocked region 1 kyr post-explosion (right side, panel a). Later, the ratio remains above unity only at the reflected shock front and in cool clumps/filaments formed after reverberation (right side, panels b and c). With $S_{\rm max} = 10^8$ dyn cm$^{-2}$ and $v_d = 500$ km s$^{-1}$ (lower panels), $\Omega_{\rm rMET}$ also surpasses $\omega_{cri}$ at the shock front and in localized cooling regions post-reverberation. Nevertheless, as discussed by \citet{HoangandLee2020}, dust grains immersed in hot gas likely have compact structures with high tensile strengths (see equation \ref{eq:smax}).

\section{Radiative Torque Disruption}
\label{app:RATD}

Similarly to the disruption of dust grains by mechanical torques, if the dust grains are exposed to an intense radiation field, the radiation can rapidly spin them up, possibly leading to their disruption once they exceed the maximum tensile strength (referred to as Radiative Torque Disruption, RATD) on a timescale \citep{Hoangetal2019}:

\begin{eqnarray}
t_{\rm RATD} &\simeq& 10^{3} (\gamma U)^{-1} \bar{\lambda}_{0.5}^{1.7} \hat{\rho}_{gr}^{1/2} S_{\max,9}^{1/2} a_{-5}^{-0.7} \, \mbox{yr}
\label{eq:RATD}
\end{eqnarray}

for $s^{1/3} a \lesssim \bar{\lambda}/2.5$, where $\gamma \approx 1$ for a unidirectional radiation field, $U$ is the radiation field strength normalized to the interstellar radiation field, and $\hat{\rho}_{gr} = \rho_{\rm gr}/3$ g cm$^{-3}$ is the normalized grain density. $S_{\max,9} = S_{\max}/10^{9}$ dyn cm$^{-2}$ is the maximum tensile strength normalized to $10^{9}$ dyn cm$^{-2}$, and $\bar{\lambda}_{0.5} = \bar{\lambda}/0.5\,\mu$m is the mean wavelength of the radiation field scaled to 0.5 $\mu$m.

The maximum angular velocity of dust grains spun-up by radiative torques, 
$\Omega_{\rm RAT}$, is given by the equation \citep{HoangandTram2022}

\begin{eqnarray}
\Omega_{\rm RAT} 
&\simeq& 9.4 \times 10^{7} s^{1/3} a_{-5} \left( \frac{\bar{\lambda}}{1.2 \, \mu \mathrm{m}} \right)^{-2} \left( \frac{\gamma U_{6}}{n_{8} T_{\rm gas, 1}^{1/2}} \right) \left( \frac{1}{1 + F_{\rm IR}} \right) \, \mathrm{rad} \, \mathrm{s}^{-1},
\end{eqnarray}

where $\gamma$ is the anisotropy degree of the radiation field, $U$ is the energy density of the radiation field, and $F_{\rm IR}$ is the infrared damping factor, given by $F_{\rm IR} = 3.81 \times 10^{-3} s^{-1/3} a_{-5}^{-1} U_6^{2/3} n_8^{-1} T_{\rm gas,1}^{-1/2}$, where $U_6 = U / 10^6$, $n_8 = n_H / 10^8$ cm$^{-3}$, and $T_{\rm gas,1} = T_{\rm gas} / 10$ K.

The RATD mechanism has been shown to operate on dust grains already present in the close vicinity to  supernova explosions \citep{Hoangetal2019}. Using parameters $U = 10^6$, $\gamma = 1$ (unidirectional radiation field), $\bar{\lambda} = 0.28 \mu$m, appropriate for type II-P SNe \citep{Hoangetal2019}, and parameters from this study $n_H \sim 10^{-2}$ cm$^{-3}$, $T_{\rm gas} \sim 10^8$ K, $s = 1/2$, and $S_{\max} = 10^9$ dyn cm$^{-2}$, the size at which $\Omega_{\rm RAT} = \omega_{\rm CRI}$ (with a value of $4.2 \times 10^{10}$ rad/s) is approximately 0.014 $\mu$m. For pre-existing dust grains larger than this size, the RATD timescale is $t_{\rm RATD}\lesssim 93$ days, while smaller grains will not be torn apart by the RATD mechanism. As dust formation in the ejecta of core-collapse supernovae is expected to commence at hundreds of days after the supernova explosion \citep{Sluderetal2018}, the RATD mechanism will not be acting on such ejecta dust grains.

In the present study, the focus has been on wind-blown bubbles formed by single massive stars, while the effect of radiation from external sources, such as nearby star clusters, has not been included. Nevertheless, as shown in \citet{MartinezGonzalezetal2014} and \citet{MartinezGonzalezetal2021}, the surrounding wind-driven shell acts as a protective barrier, preventing both ionizing and non-ionizing radiation from reaching the bubble's interior. Thus, one can anticipate a minor effect of nearby star clusters on the dust evolution in supernova remnants evolving within wind-blown bubbles.

\vspace{0.4cm}

\bibliographystyle{apj}
\bibliography{master.bib}

\end{document}